\documentclass[pra,twocolumn,superscriptaddress]{revtex4-1}
\usepackage{float,graphicx,multirow, amsmath, color,dsfont,amsthm, amssymb,comment}
\usepackage{physics}
\newtheorem{prop}{Proposition}
\usepackage{xcolor}
\definecolor{maroon}{RGB}{100,20,20}
\definecolor{dblue}{RGB}{20,20,100}
\usepackage[colorlinks=true,linkcolor=dblue,citecolor=maroon,urlcolor=maroon]{hyperref}

\begin{document}
\title{
Revealing quantum contextuality using a single 
measurement device
}
\author{Jaskaran Singh}
\email{jaskaran@us.es}
\affiliation{Departamento de f\'{\i}sica Aplicada II, Universidad de Sevilla, E-41012 Sevilla, Spain}
\affiliation{Department of Physical Sciences, Indian
Institute of Science Education and Research (IISER) Mohali,
Sector 81 SAS Nagar, Manauli PO 140306 Punjab India.}
\author{Rajendra Singh Bhati}
\email{ph16076@iisermohali.ac.in}
\affiliation{Department of Physical Sciences, Indian
Institute of Science Education and Research (IISER) Mohali,
Sector 81 SAS Nagar, Manauli PO 140306 Punjab India.}
\author{Arvind}
\email{arvind@iisermohali.ac.in}
\affiliation{Department of Physical Sciences, Indian
Institute of Science Education and Research (IISER) Mohali,
Sector 81 SAS Nagar, Manauli PO 140306 Punjab India.}
\affiliation{Punjabi University Patiala,
147002, Punjab, India}
\begin{abstract}
In this work we analyse the notion of measurement
non-contextuality (MNC) and identify contextual scenarios
which involve sequential measurements of only a single
measurement device.  We show that any non-contextual
ontological model fails to explain the statistics of
outcomes of a single carefully constructed positive operator
valued measure (POVM) executed sequentially on a quantum
system. The context of measurement arises from the different
configurations in which the device can be used. We develop
an inequality from the non-contextual (NC) ontic model, and
construct a quantum situation involving measurements from
the KCBS inequality. We show that the resultant statistics
arising from this device violate our NC inequality.  This
device can be generalised by incorporating measurements from
arbitrary $n$-cycle contextuality inequalities of which $n =
5$ corresponds to the KCBS inequality. We show that the NC
and quantum bounds for various scenarios 
can be derived more easily using only the functional
relationships between the outcomes for larger values
$n$. This makes it one of the
simpler contextual inequalities to analyse.
\end{abstract}
\maketitle
\section{Introduction} Quantum theory is contextual since
the outcomes of measurements depend on the measurement
context, namely, the set of commuting observables being
measured along with the desired observable~\cite{Koc,lsw}.
While Bell type
inequalities~\cite{bell_hidd_var,bell_non_locality,chsh} can
reveal quantumness of composite systems, quantum
contextuality can be  demonstrated on a single indivisible
system. The simplest such scenario involves a three
dimensional quantum system and  five different projective
measurements~\cite{kcbs}. While the first proof that quantum
theory is contextual was provided by Kochen and
Specker~\cite{ks_theorem}, over time a number of simpler and
more systematic ways of revealing quantum contextuality,
particularly the ones based on graph theory, have become
available~\cite{graph_cabello,acin2015}.  The graph
theoretic approach has been successful in identifying new
contextual
scenarios~\cite{nc_n_cycle,e_principle_corr,context_dago,context_seven_set,
necessary_cond_context,optimal_ineq_context,
cabello_13_ray,oh_13_ray,cabello_twin_ineq,kishor_context_ncycle},
simplifying formulations of contextuality
monogamy~\cite{context_mono}, contextuality non-locality
relationship~\cite{context_nonlocal_mono},  developing
robust self tests~\cite{kishor_robust} and information
theoretic applications of quantum
contextuality~\cite{jask_mono,cabello_context_keydist,troupe_context_keydist,
debasish2}.

The approach developed by Spekkens \textit{et al.}~\cite{spekkens_context, spekkens1, schmid3} provides
additional insights. Since its inception, this approach has
received tremendous research attention and has been
influential in exhibiting advantage in various information
theoretic applications like parity oblivious bit
transfer~\cite{spekkens_parity_context,context_corr_quantum},
random access codes~\cite{rac_context} state
discrimination~\cite{schmid_state_discrimination_context}
and quantum key distribution~\cite{anubhav,jask_mono}.
Spekkens' approach generalizes the notion of contextuality
to positive-operator-valued measures (POVMs), as well as
provides a notion of contextuality for preparations and
transformations.  This generalized approach has facilitated
research in other aspects of foundations of quantum
theory~\cite{pusey, pusey2}.  Furthermore, this
generalization provides a technique for noise-robust
experimental demonstration  of
contextuality~\cite{kunjwal_noncontext_unphys,kunjwal_stat_robust_context,kunjwal_kstheorem_no_det,debasish1,anwer}
and leads to information theoretic applications of quantum
situations involving preparation
contextuality~\cite{spekkens_prep_context_app,sikora_random_access_codes,banik_context_app}.
Using this approach the minimum number of measurements
required to reveal quantum contextuality so far is
three~\cite{kunjwal_context_qubit}, while no physical
principle prohibits a smaller number. So far a full
characterization of minimum number of measurement devices
required to exhibit quantum contextual correlations has been
achieved for projective measurements, however, for POVMs the
scenario is completely different, non-trivial and an open
research problem.  Recently it was shown in
Ref.~\cite{schmid} that it is possible to talk about quantum
contextuality for only a single measurement device by using
the property of convexity of ontic distributions and
response functions~\cite{spekkens1}.

In this work we analyse how coarse graining of measurement
outcomes can help reveal quantum contextuality for even a
single measurement device. These coarse grainings and their
representation in ontic models have been well studied in
Ref.~\cite{spekkens1}.  We formulate a new experimentally
verifiable NC inequality for coarse grained measurement
outcomes which exhibits a violation for a single carefully
constructed measurement device, which is a positive operator
valued measure (POVM).  The POVM is executed twice
sequentially to reveal the contextuality of the quantum
situation involved. The POVM itself is constructed by coarse
graining over the projective measurements that appear in a
test for revealing quantum contextuality. While we
explicitly show an example using the measurements that
appear in the KCBS inequality, our result readily
generalises to the case of $n$-cycle contextuality
inequalities.

The inequality that we present is independent of any
underlying quantum state and only depends on outcomes of the
measuring device. This novel feature of our inequality
clearly brings out the nature of sequential measurements.
While other similar inequalities~\cite{selby_context}
evaluate joint probability distributions over two or more
measurements and indirectly imply conclusions on sequential
measurements, we take a more direct approach and evaluate
conditional probabilities instead, which provides
straightforward conclusions on the sequential measurements
themselves. Furthermore, our result can also
incorporate more scenarios based on multiple sequential measurements.
The corresponding NC and quantum bounds can be calculated
following the prescription we provide for a single
measurement device applied sequentially.

The paper is organized as follows. In Sec.~\ref{sec:sec1}, we briefly review the assumptions of 
measurement and preparation non-contextuality. In Sec.~\ref{sec:sec2} we describe our setup of the POVM
to be measured which will be used as a signature of contextuality. In Sec.~\ref{sec:sec3} we derive a NC inequality based on the assumptions of measurement and preparation NC. In this section we also explicitly derive the quantum bound and show a violation of the inequality. In Sec.~\ref{sec:sec4} we provide some conclusive remarks on our result.
	
\section{Measurement and preparation non-contextuality }
\label{sec:sec1}

The definition of measurement NC as formulated by
Spekkens~\cite{spekkens_context}, relies on defining a
notion of equivalence among different experimental procedures.
Specifically, two measurement procedures $\mathcal{M}$ and
$\mathcal{M}'$ are deemed equivalent if they yield the same
statistics for every possible preparation procedure $\text{P}$, that
is
\begin{equation}
p(k|\mathcal{M},\text{P})=p(k|\mathcal{M}',\text{P})
\implies \left[k|\mathcal{M}\right]=\left[k|\mathcal{M}'\right],
\label{eq:meas_equiv}
\end{equation}
where $p(k|\mathcal{M},\text{P})$ is the
probability of obtaining the outcome $k$ in the measurement
$\mathcal{M}$ for preparation $\text{P}$ and $\left[k|\mathcal{M}\right]$ 
denotes the event of obtaining the outcome $k$ in a measurement $\mathcal{M}$.
Consider an ontic model where an
ontic state $\lambda$ can be used to predict the measurement
outcomes.
Based on the
definition of equivalence classes for measurement
procedures, and  as motivated
by Leibniz's principle of indiscernibles, the
definition of measurement NC  is
\begin{equation}
\begin{aligned}
&p(k|\mathcal{M},\text{P})=p(k|\mathcal{M}',\text{P}) ~\forall \text{P}\\
&\implies
\xi(k|\lambda,\mathcal{M})=\xi(k|\lambda,\mathcal{M}^\prime) ~\forall \lambda,
\end{aligned}
\label{eq:meas_noncontext}
\end{equation}
where  $\xi(k|\lambda,\mathcal{M})$ is the  epistemic
response function for
assigning the probability of obtaining an outcome $k$ in a
measurement $\mathcal{M}$ given the ontic state $\lambda$
with
\begin{equation}
\xi(k|\lambda,\mathcal{M})\leq 1, \quad
\sum_k\xi(k|\lambda,\mathcal{M}) = 1 ~\forall \lambda. 
\end{equation}

Thus, the assumption of measurement NC implies that the
response function for different outcomes of equivalent
measurement procedures is the same if their observed
statistics are the same for all preparation procedures.

In a similar fashion, one can also define a notion of
preparation NC. Firstly, we define two preparation
procedures $\text{P}$ and $\text{P}'$ to be equivalent if
they yield the same statistics for all possible measurement
procedures $\mathcal{M}$,
\begin{equation}
p(k|\mathcal{M}, P) = p(k|\mathcal{M}, P') \implies \left[k|P\right] = \left[k|P'\right],
\end{equation}
where $\left[k|P\right]$ denotes the event in which an outcome $k$ is obtained when the preparation was $P$.

Like measurement NC, the assumption of preparation NC is stated as
\begin{equation}
\begin{aligned}
&p(k|\mathcal{M},\text{P})=p(k|\mathcal{M},\text{P}')~\forall\mathcal{M} \\
&\implies
\mu(\lambda|\text{P}) = \mu(\lambda|\text{P}')~\forall \lambda,
\end{aligned}
\end{equation}
where $\mu(\lambda|P)$ is an epistemic distribution over the
ontic variables $\lambda$ given a preparation procedure
$\text{P}$, with
\begin{equation}
\int_{\lambda\in \Lambda}d\lambda \mu(\lambda|\text{P}) = 1,
\end{equation}
where $\Lambda$ is the space of all ontic variables $\lambda$. 
Thus, two preparation procedures are assigned the same epistemic state if the 
observed statistics for all measurements applied on them are the same.

\section{The setup} 
\label{sec:sec2}

Here we explicitly construct a non-trivial scenario on which
we can apply the assumption of measurement NC with
post-processing of measurement outcomes. While the scenario
leads to the
exhibition of measurement NC using a single measurement
device, it also serves as a stepping point in the
prescription NC for arbitrary
sequential measurement devices.

Consider a measurement procedure able to function in two
different configurations and we assume a non-contextual
ontological model explaining the statistics of both the
configurations.
\begin{figure}
\includegraphics[scale=1]{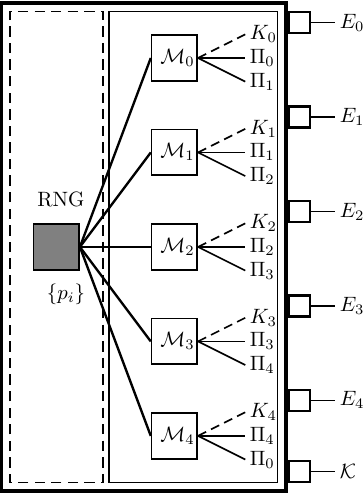}
\caption{A schematic diagram of the two configurations of
the measurement device $\mathcal{M}$. In the first
configuration, the device samples the measurements
$\mathcal{M}_i$ from a random number generator (RNG) with
probability $p_i$. In this configuration the dashed outcomes are 
collected as final outcome $\mathcal{K}$, while outcomes with solid 
lines are collected as final outcomes $E_i$. In the second configuration, each
measurement $\mathcal{M}_i$ can be performed independently
with outcomes given by solid and 
dashed lines.}
\label{fig:setup}
\end{figure}
The schematics of our setup is shown in Fig~\ref{fig:setup}.
The setup is comprised of $5$ projective measurements
$\lbrace \mathcal{M}_i\rbrace$, with three outcomes each. We
choose these projective measurements to be the ones utilized
in the KCBS scenario. We also assume that the projectors involved in each measurement 
are of rank-one. Using these measurements we can enable
the device to work in two possible configurations as
detailed below:

\begin{itemize}
\item[C1:]In this configuration the measurements
$\lbrace\mathcal{M}_i\rbrace$ are 
sampled from a probability distribution $\lbrace
p_i\rbrace$, $i\in\lbrace0,1,2,3,4\rbrace$, and the device
implements the projective 
measurement $\lbrace\Pi_i, \Pi_{i\oplus 1}, K_i\rbrace$,
where $K_i$ is added to complete the measurement
and $\oplus$  is addition modulo $5$. The projectors 
$\lbrace \Pi_i\rbrace$
satisfy 
${\rm Tr}(\Pi_i\Pi_{i\oplus 1})=0$ and are 
the same as the ones used in the derivation of KCBS
inequality. The resultant measurement is then a POVM
$\mathcal{M}$
with outcomes,
\begin{equation}
\mathcal{M}: \lbrace E_0,E_1,E_2,E_3,E_4,\mathcal{K}\rbrace, 
\label{eq:povm}
\end{equation}
where $E_i=(p_i+p_{i\oplus 4})\Pi_i$ and $\mathcal{K} = \sum_ip_iK_i$ with 
\begin{equation}
\sum_{0}^{4} E_i=
\sum_{i=0}^{4}(p_i+p_{i\oplus 4})\Pi_i\leq \mathds{1},
\label{eq:povm_completeness}
\end{equation}
which is a consequence of completeness of measurement.

\item[C2:] In this configuration
we choose a particular setting $i$, implementing
a projective measurement 
on our device
\begin{equation}
\mathcal{M}_i: \lbrace\Pi_i,\Pi_{i\oplus 1},K_i\rbrace.
\label{eq:pvm}
\end{equation} 

This is akin to
blocking all measurement outcomes $j\neq i$. From
completeness we again have
\begin{equation}
\Pi_i + \Pi_{i \oplus 1} \leq \mathds{1} \quad \forall i.
\label{eq:pvm_completeness}
\end{equation} 

Five such settings are
possible corresponding to the
projective measurements 
detailed above and each is labeled by a value of $i$.
\end{itemize}

We now formulate an ontological description of the
aforementioned measurement device. For each measurement
outcome $\Pi_i$ from the measurement $\mathcal{M}_i$, there
corresponds a response function
$\xi(\Pi_i|\lambda,\mathcal{M}_{i})$, where $\lambda$s are
the ontic variables. Therefore an ontological description of
$\mathcal{M}_i$ will have the form
\begin{equation}
\mathcal{M}_{i,\lambda}:
\lbrace\xi(\Pi_i|\lambda,\mathcal{M}_{i}),\xi(\Pi_{i\oplus
1}|\lambda,\mathcal{M}_{i}),\xi(K_i|\lambda,\mathcal{M}_{i})\rbrace.
\end{equation}
This ontological model of configuration C2 induces an
ontological description for the configuration C1. This is
due the fact that the generalized measurement procedure
$\mathcal{M}$ can be described in terms of post-processing
of projective measurements defined for C2 configuration
\textit{i.\,e.} $\{\mathcal{M}_i|i=0,1,2,3,4\}$.  Let the
ontological model of the measurement device in C1
configuration be described as,

\begin{equation}
\mathcal{M}_\lambda: \left\{
\xi'(E|\lambda,\mathcal{M})|\,E\in\{E_0,E_1,E_2,E_3,E_4,\mathcal{K}\}
\right\},
\end{equation}

Here $\xi'(E|\lambda,\mathcal{M})$ denotes the response
function corresponding to the outcome
$E\in\{E_0,E_1,E_2,E_3,E_4,\mathcal{K}\}$ when the system is
measured in C1 configuration. Since C1 implements the
projective measurement $\mathcal{M}_i$ with probability
$p_i$, we have for $i\in\{0,1,2,3,4\}$,
\begin{equation}
\label{resp_C1}
\xi'(E_i|\lambda, \mathcal{M})=p_i\xi(\Pi_i|\lambda,
\mathcal{M}_i)+p_{i\oplus 4}\xi(\Pi_i|\lambda,
\mathcal{M}_{i\oplus 4})
\end{equation}
and
\[
\xi'(\mathcal{K}|\lambda, \mathcal{M}) =
\sum^{4}_{i=0}p_i\xi(K_i|\lambda, \mathcal{M}_i)
\]
such that
$\xi(\Pi_i|\lambda,\mathcal{M}_i)$,
$\xi'(E_i|\lambda,\mathcal{M})$,
$\xi^\prime(\mathcal{K}|\lambda,\mathcal{M})$, and
$\xi(K_i|\lambda, \mathcal{M}_i)\in\left[
0,1\right]$. It is well known that the description for
response functions presented above follows from the
probability theory in a straightforward way and have nothing
to do with assumption of measurement non-contextuality (see
Ref.~\cite{spekkens1}, for example).  Similarly, the
epistemic state of a system when it is prepared in a state
corresponding to the outcome $E_i$ of measurement procedure
$\mathcal{M}$ is described as
\begin{equation}
\label{epis_C1}
\mu^\prime(\lambda|E_i,\mathcal{M})=
p_i\mu(\lambda|\Pi_i,\mathcal{M}_i)+p_{i\oplus{4}}\mu(\lambda|\Pi_i,\mathcal{M}_{i\oplus{4}})
\end{equation}
where $\mu(\lambda|\Pi_i,\mathcal{M}_j)$, for $j=i,i\oplus4$, is the epistemic state
of the ontological variable when the corresponding system is
prepared in $\Pi_i$ using measurement $\mathcal{M}_j$ in C2 configuration.
Note that, just like Eq.~\eqref{resp_C1}, Eq.~\eqref{epis_C1} also follows
from the probability theory and has nothing
to do with the assumption of preparation non-contextuality.

\section{Contextuality from a single measurement device}
\label{sec:sec3}

We are now ready to propose an inequality to be tested by
sequential measurements and explicitly derive its maximum
non-contextual value. We then show that a quantum
description of the same leads to a violation. While we focus
on the measurement device set up in Fig.~\ref{fig:setup},
the technique that we use to derive the NC and quantum bound
can be easily modified for arbitrary measurement devices.

The scenario is as follows. The setup with
configuration C1 described in Sec.~III is used to perform a
measurement on a state $\rho$ which is chosen such tr$(\rho
\Pi_i) \neq 0$ $\forall \,\,i$. Two cases arise: (i) if an
outcome $E_i$ is observed, the setup with configuration C1
is again used to perform a second measurement on the
resultant post-measurement state corresponding to the outcome
$E_i$, and (ii) in case the outcome $\mathcal{K}$ is
observed in the first measurement, the corresponding
experimental run is discarded.  For case (i) the following
proposition holds true.

\begin{prop}
Consider $\mathcal{C}=\sum_{i,j=0}^4p(E_j|E_i)$, where
$p(E_j|E_i)$
denotes the conditional probability of
obtaining outcome $E_i$ given $E_j$ when measurement
$\mathcal{M}$ is performed twice sequentially in C1
configuration. Then, the following relation holds if we assume
$p_i=s,\,\,\forall\, i$,
\[
\mathcal{C}_{QM}=\frac{1}{2s}\mathcal{C}_{NC},
\]
where $\mathcal{C}_{QM}$ and $\mathcal{C}_{NC}$ are
values of $\mathcal{C}$ obtained using quantum theory
and non-contextual ontological models respectively.
\end{prop}	

\begin{proof}
Consider an experimental scenario in which a system is
prepared in a state that corresponds to an outcome $a$ of a
measurement $\mathcal{A}$. The system then undergoes a
measurement denoted by $\mathcal{B}$ with an outcome denoted
by $b$. Let us represent the epistemic state and the
response function for this scenario as
$\mu(\lambda|a,\mathcal{A})$ and
$\xi(b|\lambda,\mathcal{B})$ respectively. Then the
probability of obtaining an outcome $b$ in a measurement
$\mathcal{B}$ such that the state was prepared by the
outcome $a$ of measurement $\mathcal{A}$ is given by
\begin{equation}
p(b|a,\mathcal{A},\mathcal{B})=
\int_{\lambda\in\Lambda}\xi(b|\lambda,\mathcal{B})\mu(\lambda|a,\mathcal{A})d\lambda.
\label{eq:prep_meas}
\end{equation}
	
Let us consider that the measurement $\mathcal{M}$ was
performed twice sequentially in C1 configuration playing
role of preparation $\mathcal{A}$ and the subsequent
measurement $\mathcal{B}$.  Therefore in our case we have
$\mathcal{A} = \mathcal{B} = \mathcal{M}$. Following
Eq.~\eqref{eq:prep_meas}, the conditional probability of
obtaining the outcome $E_j$ given that the state was
prepared by the outcome $E_i$ of the measurement
$\mathcal{M}$ is given by   
	
\begin{equation}
p(E_j|E_i,\mathcal{M},\mathcal{M})=
\int_{\lambda\in\Lambda}\xi^\prime(E_j|\lambda,\mathcal{M})\mu^\prime(\lambda|E_i,\mathcal{M})d\lambda.
\end{equation}
Since the measurement setting is specified in our
setup and remains unchanged throughout the implementation,
we will use short notation $p(E_j|E_i)$ instead of
$p(E_j|E_i,\mathcal{M},\mathcal{M})$ hereafter.

Since for the configuration C2, we have
$p(\Pi_i|\mathcal{M}_i,P)=p(\Pi_i|\mathcal{M}_{i\oplus
4},P)$ for all preparations $P$ at the operational level,
the assumption of measurement non-contextuality asserts
	
\begin{equation} \label{meas_nc_i}
\xi(\Pi_j|\lambda, \mathcal{M}_i)=\xi(\Pi_j|\lambda,
\mathcal{M}_{j\oplus 4}).  \end{equation}

Similarly, for C2 configuration, the assumption of
preparation non-contextuality implies

\begin{equation}
\label{prep_nc_i}
\mu(\lambda|\Pi_i,\mathcal{M}_i)=\mu(\lambda|\Pi_{i},\mathcal{M}_{i\oplus 4}).
\end{equation}

Note that the two measurement settings $\mathcal{M}_i$ and
$\mathcal{M}_{i\oplus 4}$ define two different contexts for
the same preparation state $\Pi_i$ here.  Using these
results and the observation of Ref.~\cite{spekkens1}, Eqs.
\eqref{resp_C1} and \eqref{epis_C1} can be re-written as
as,
	
\begin{equation}
\begin{aligned}
\xi^\prime(E_j|\lambda,\mathcal{M})&=
(p_j+p_{j\oplus 4})\xi(\Pi_j|\lambda,\mathcal{M}_j),  \\
\mu^\prime(\lambda|E_{i},\mathcal{M})&=
(p_i+p_{i\oplus 4})\mu(\lambda|\Pi_{i},\mathcal{M}_i),
\end{aligned}
\end{equation}
which gives us,
\begin{equation}
\begin{aligned}
&p(E_j|E_i)=\int_{\lambda\in\Lambda}\xi^\prime(E_j|\lambda,\mathcal{M})
\mu^\prime(\lambda|E_i,\mathcal{M})d\lambda
\\
&=
(p_i+p_{i\oplus{4}})(p_j+p_{j\oplus{4}})
p(\Pi_j|\Pi_i,\mathcal{M}_i,\mathcal{M}_j),
\end{aligned}
\end{equation}
where,
\begin{equation}
\begin{aligned}
p(\Pi_j|\Pi_i,\mathcal{M}_i,\mathcal{M}_j)&=
\int_{\lambda\in\Lambda}\xi(\Pi_j|\lambda,\mathcal{M}_j)
\mu(\lambda|\Pi_i,\mathcal{M}_i)d\lambda.
\end{aligned}
\end{equation}
	
The fact that a valid ontological model must reproduce all
statistics of an operational theory implies
\begin{equation}
\int_{\lambda\in\Lambda}\xi(\Pi_j|
\lambda,\mathcal{M}_j)\mu(\lambda|\Pi_i,\mathcal{M}_i)d\lambda=p(\Pi_j|\Pi_i)
\end{equation}
Therefore, we have,
\begin{equation}
\mathcal{C}_{NC}=4s^2\sum_{i,j=0}^{4}p(\Pi_j|\Pi_i).
\end{equation}
	
An evaluation of the expression $\mathcal{C}$ for quantum
states and measurements can
be made in a straightforward manner using only the
functional constraints between $E_j$ and
$\Pi_i$. We have for quantum theory,
\begin{equation}
\begin{aligned}
&p(E_j|E_i)=\tr(E_j\Pi_i)=(p_j+p_{j\oplus 4})\tr(\Pi_j\Pi_i),\\
&\Rightarrow
\mathcal{C}_{QM}=2s\sum_{i,j=0}^{4}\tr(\Pi_j\Pi_i) =
2s\sum_{i, j = 0}^{4}p(\Pi_j|\Pi_i),
\end{aligned}
\end{equation}
where for the first equality we have used the fact that the
state prepared after the first measurement is $\Pi_i$.
Therefore, it can be seen that we have $\mathcal{C}_{QM}>\mathcal{C}_{NC}$.
More specifically we have, 
\begin{equation}
\mathcal{C}_{QM} = \frac{1}{2s}\mathcal{C}_{NC},
\label{eq:final}
\end{equation}
which concludes the proof. 
\end{proof}

It should also be noted that while quantum probabilities are
given by the Born rule, we make no such claim on the
probabilities derived under a non-contextual assumption.
Instead, we only assume that the functional constraints on
the probabilities for both the scenarios must be satisfied
at all times. In this way we are able to derive
Eq.~\eqref{eq:final} by using the fact that NC conditional
probabilities must give the same predictions as the quantum
mechanics
for each sequential measurement independently.

It should be noted that the relation between
$\mathcal{C}_{QM}$ and $\mathcal{C}_{NC}$ is established by
alluding to a particular set of quantum states $\rho$ on
which the first measurement is performed.  Specifically, we
require that the initial state $\rho$ should not be
orthogonal to any of the projectors $\Pi_i$. This is to
ensure that $p(E_i) \neq 0$ $\forall \,\,i$ in the first
measurement so that the
quantities $p(E_j|E_i)$ are well defined. Therefore, apart
from the aforementioned set of states, Proposition $1$ holds
true for all states.

It should also be
noted that while no quantum measurements have been referred
to, the orthogonality relations between them have been
specified as well as the functional constraints between the
outcomes of the measurement device in the two
configurations.

This simplified signature of measurement NC also allows for
easily incorporating arbitrary $n$-cycle contextuality
scenarios ($n\geq 5$)~\cite{nc_n_cycle} like as we did for the KCBS scenario. For an
arbitrary $n$-cycle scenario instead of KCBS in the
measurement device described above, we have,
\begin{equation}
\begin{aligned}
\mathcal{C}_{NC} &= 4s^2\sum^{n-1}_{i,j=0}p(\Pi_j|\Pi_i), \\
\mathcal{C}_{QM} &= 2s\sum^{n-1}_{i,j =0}\text{tr}(\Pi_j\Pi_i),
\end{aligned}
\end{equation}
where in the second equality we have used the standard Born rule, while in the first equality we only 
assume that the probability distributions are assigned in a non-contextual manner as explained above.

As can be seen the functional relationship between the
quantum and contextual bound retains the same form. However,
for increasing $n$, $p_i = s$ decreases, which consequently
broadens the gap between the contextual and non-contextual
bound.

\section{Conclusion} 
\label{sec:sec4}

In this work we show that the notion of preparation and
measurement NC can lead to several new  and interesting
scenarios which exhibit a superiority of
quantum over classical correlations by violating a
novel NC bound on an ontic model for a single measurement
device.  We have detailed a scenario in which the statistics
of outcomes from a single measurement device is not
reproducible from an underlying non-contextual ontic model. 
 The inequality that we
propose is based on conditional probabilities of outcomes of
two (or more) measurements. This way we can directly make
implications on the (non-)contextual nature of sequential
measurements. Our inequality can be generalized for
arbitrary sequential measurements for which the NC bound can
be calculated easily using our prescription.

To the best of our knowledge such a scenario has been
constructed for the first time. Our results pave the way
for future theoretical as well as experimental work to
unearth the contextuality of arbitrary sequential quantum
measurements. It would be interesting to find out whether
the inequality we propose is optimal in the number of
measurements and/or outcomes.
Another extension of our work would entail generalization of
the single measurement device scenario to incorporate
multiple sequential measurements of the same or different
non-contextual scenarios involving $n$-outcomes which may
not be cyclic.

{\it Acknowledgements.---}
JS is supported by QuantERA grant SECRET, by
\href{http://dx.doi.org/10.13039/501100011033:}{MCINN/AEI}(Project
No.\ PCI2019-111885-2)
A  acknowledge the financial
support from {\sf DST/ICPS/QuST/Theme-1/2019/General}
Project number {\sf Q-68}.

\begin{thebibliography}{48}%
	\makeatletter
	\providecommand \@ifxundefined [1]{%
		\@ifx{#1\undefined}
	}%
	\providecommand \@ifnum [1]{%
		\ifnum #1\expandafter \@firstoftwo
		\else \expandafter \@secondoftwo
		\fi
	}%
	\providecommand \@ifx [1]{%
		\ifx #1\expandafter \@firstoftwo
		\else \expandafter \@secondoftwo
		\fi
	}%
	\providecommand \natexlab [1]{#1}%
	\providecommand \enquote  [1]{``#1''}%
	\providecommand \bibnamefont  [1]{#1}%
	\providecommand \bibfnamefont [1]{#1}%
	\providecommand \citenamefont [1]{#1}%
	\providecommand \href@noop [0]{\@secondoftwo}%
	\providecommand \href [0]{\begingroup \@sanitize@url \@href}%
	\providecommand \@href[1]{\@@startlink{#1}\@@href}%
	\providecommand \@@href[1]{\endgroup#1\@@endlink}%
	\providecommand \@sanitize@url [0]{\catcode `\\12\catcode `\$12\catcode
		`\&12\catcode `\#12\catcode `\^12\catcode `\_12\catcode `\%12\relax}%
	\providecommand \@@startlink[1]{}%
	\providecommand \@@endlink[0]{}%
	\providecommand \url  [0]{\begingroup\@sanitize@url \@url }%
	\providecommand \@url [1]{\endgroup\@href {#1}{\urlprefix }}%
	\providecommand \urlprefix  [0]{URL }%
	\providecommand \Eprint [0]{\href }%
	\providecommand \doibase [0]{http://dx.doi.org/}%
	\providecommand \selectlanguage [0]{\@gobble}%
	\providecommand \bibinfo  [0]{\@secondoftwo}%
	\providecommand \bibfield  [0]{\@secondoftwo}%
	\providecommand \translation [1]{[#1]}%
	\providecommand \BibitemOpen [0]{}%
	\providecommand \bibitemStop [0]{}%
	\providecommand \bibitemNoStop [0]{.\EOS\space}%
	\providecommand \EOS [0]{\spacefactor3000\relax}%
	\providecommand \BibitemShut  [1]{\csname bibitem#1\endcsname}%
	\let\auto@bib@innerbib\@empty
	\bibitem [{\citenamefont {Simon~Kochen}(1968)}]{Koc}%
	\BibitemOpen
	\bibfield  {author} {\bibinfo {author} {\bibfnamefont {E.~S.}\ \bibnamefont
			{Simon~Kochen}},\ }\href@noop {} {\bibfield  {journal} {\bibinfo  {journal}
			{Indiana Univ. Math. J.}\ }\textbf {\bibinfo {volume} {17}},\ \bibinfo
		{pages} {59} (\bibinfo {year} {1968})}\BibitemShut {NoStop}%
	\bibitem [{\citenamefont {Liang}\ \emph {et~al.}(2011)\citenamefont {Liang},
		\citenamefont {Spekkens},\ and\ \citenamefont {Wiseman}}]{lsw}%
	\BibitemOpen
	\bibfield  {author} {\bibinfo {author} {\bibfnamefont {Y.-C.}\ \bibnamefont
			{Liang}}, \bibinfo {author} {\bibfnamefont {R.~W.}\ \bibnamefont {Spekkens}},
		\ and\ \bibinfo {author} {\bibfnamefont {H.~M.}\ \bibnamefont {Wiseman}},\
	}\href {\doibase https://doi.org/10.1016/j.physrep.2011.05.001} {\bibfield
		{journal} {\bibinfo  {journal} {Physics Reports}\ }\textbf {\bibinfo {volume}
			{506}},\ \bibinfo {pages} {1 } (\bibinfo {year} {2011})}\BibitemShut
	{NoStop}%
	\bibitem [{\citenamefont {Bell}(1966)}]{bell_hidd_var}%
	\BibitemOpen
	\bibfield  {author} {\bibinfo {author} {\bibfnamefont {J.~S.}\ \bibnamefont
			{Bell}},\ }\href {\doibase 10.1103/RevModPhys.38.447} {\bibfield  {journal}
		{\bibinfo  {journal} {Rev. Mod. Phys.}\ }\textbf {\bibinfo {volume} {38}},\
		\bibinfo {pages} {447} (\bibinfo {year} {1966})}\BibitemShut {NoStop}%
	\bibitem [{\citenamefont {Brunner}\ \emph {et~al.}(2014)\citenamefont
		{Brunner}, \citenamefont {Cavalcanti}, \citenamefont {Pironio}, \citenamefont
		{Scarani},\ and\ \citenamefont {Wehner}}]{bell_non_locality}%
	\BibitemOpen
	\bibfield  {author} {\bibinfo {author} {\bibfnamefont {N.}~\bibnamefont
			{Brunner}}, \bibinfo {author} {\bibfnamefont {D.}~\bibnamefont {Cavalcanti}},
		\bibinfo {author} {\bibfnamefont {S.}~\bibnamefont {Pironio}}, \bibinfo
		{author} {\bibfnamefont {V.}~\bibnamefont {Scarani}}, \ and\ \bibinfo
		{author} {\bibfnamefont {S.}~\bibnamefont {Wehner}},\ }\href {\doibase
		10.1103/RevModPhys.86.419} {\bibfield  {journal} {\bibinfo  {journal} {Rev.
				Mod. Phys.}\ }\textbf {\bibinfo {volume} {86}},\ \bibinfo {pages} {419}
		(\bibinfo {year} {2014})}\BibitemShut {NoStop}%
	\bibitem [{\citenamefont {Clauser}\ \emph {et~al.}(1969)\citenamefont
		{Clauser}, \citenamefont {Horne}, \citenamefont {Shimony},\ and\
		\citenamefont {Holt}}]{chsh}%
	\BibitemOpen
	\bibfield  {author} {\bibinfo {author} {\bibfnamefont {J.~F.}\ \bibnamefont
			{Clauser}}, \bibinfo {author} {\bibfnamefont {M.~A.}\ \bibnamefont {Horne}},
		\bibinfo {author} {\bibfnamefont {A.}~\bibnamefont {Shimony}}, \ and\
		\bibinfo {author} {\bibfnamefont {R.~A.}\ \bibnamefont {Holt}},\ }\href
	{\doibase 10.1103/PhysRevLett.23.880} {\bibfield  {journal} {\bibinfo
			{journal} {Phys. Rev. Lett.}\ }\textbf {\bibinfo {volume} {23}},\ \bibinfo
		{pages} {880} (\bibinfo {year} {1969})}\BibitemShut {NoStop}%
	\bibitem [{\citenamefont {Klyachko}\ \emph {et~al.}(2008)\citenamefont
		{Klyachko}, \citenamefont {Can}, \citenamefont
		{Binicio\ifmmode~\breve{g}\else \u{g}\fi{}lu},\ and\ \citenamefont
		{Shumovsky}}]{kcbs}%
	\BibitemOpen
	\bibfield  {author} {\bibinfo {author} {\bibfnamefont {A.~A.}\ \bibnamefont
			{Klyachko}}, \bibinfo {author} {\bibfnamefont {M.~A.}\ \bibnamefont {Can}},
		\bibinfo {author} {\bibfnamefont {S.}~\bibnamefont
			{Binicio\ifmmode~\breve{g}\else \u{g}\fi{}lu}}, \ and\ \bibinfo {author}
		{\bibfnamefont {A.~S.}\ \bibnamefont {Shumovsky}},\ }\href {\doibase
		10.1103/PhysRevLett.101.020403} {\bibfield  {journal} {\bibinfo  {journal}
			{Phys. Rev. Lett.}\ }\textbf {\bibinfo {volume} {101}},\ \bibinfo {pages}
		{020403} (\bibinfo {year} {2008})}\BibitemShut {NoStop}%
	\bibitem [{\citenamefont {Kochen}\ and\ \citenamefont
		{Specker}(1975)}]{ks_theorem}%
	\BibitemOpen
	\bibfield  {author} {\bibinfo {author} {\bibfnamefont {S.}~\bibnamefont
			{Kochen}}\ and\ \bibinfo {author} {\bibfnamefont {E.~P.}\ \bibnamefont
			{Specker}},\ }\enquote {\bibinfo {title} {The problem of hidden variables in
			quantum mechanics},}\ in\ \href {\doibase 10.1007/978-94-010-1795-4_17}
	{\emph {\bibinfo {booktitle} {The Logico-Algebraic Approach to Quantum
				Mechanics: Volume I: Historical Evolution}}},\ \bibinfo {editor} {edited by\
		\bibinfo {editor} {\bibfnamefont {C.~A.}\ \bibnamefont {Hooker}}}\ (\bibinfo
	{publisher} {Springer Netherlands},\ \bibinfo {address} {Dordrecht},\
	\bibinfo {year} {1975})\ pp.\ \bibinfo {pages} {293--328}\BibitemShut
	{NoStop}%
	\bibitem [{\citenamefont {Cabello}\ \emph {et~al.}(2014)\citenamefont
		{Cabello}, \citenamefont {Severini},\ and\ \citenamefont
		{Winter}}]{graph_cabello}%
	\BibitemOpen
	\bibfield  {author} {\bibinfo {author} {\bibfnamefont {A.}~\bibnamefont
			{Cabello}}, \bibinfo {author} {\bibfnamefont {S.}~\bibnamefont {Severini}}, \
		and\ \bibinfo {author} {\bibfnamefont {A.}~\bibnamefont {Winter}},\ }\href
	{\doibase 10.1103/PhysRevLett.112.040401} {\bibfield  {journal} {\bibinfo
			{journal} {Phys. Rev. Lett.}\ }\textbf {\bibinfo {volume} {112}},\ \bibinfo
		{pages} {040401} (\bibinfo {year} {2014})}\BibitemShut {NoStop}%
	\bibitem [{\citenamefont {Ac{\'i}n}\ \emph {et~al.}(2015)\citenamefont
		{Ac{\'i}n}, \citenamefont {Fritz}, \citenamefont {Leverrier},\ and\
		\citenamefont {Sainz}}]{acin2015}%
	\BibitemOpen
	\bibfield  {author} {\bibinfo {author} {\bibfnamefont {A.}~\bibnamefont
			{Ac{\'i}n}}, \bibinfo {author} {\bibfnamefont {T.}~\bibnamefont {Fritz}},
		\bibinfo {author} {\bibfnamefont {A.}~\bibnamefont {Leverrier}}, \ and\
		\bibinfo {author} {\bibfnamefont {A.~B.}\ \bibnamefont {Sainz}},\ }\href
	{\doibase 10.1007/s00220-014-2260-1} {\bibfield  {journal} {\bibinfo
			{journal} {Communications in Mathematical Physics}\ }\textbf {\bibinfo
			{volume} {334}},\ \bibinfo {pages} {533} (\bibinfo {year}
		{2015})}\BibitemShut {NoStop}%
	\bibitem [{\citenamefont {Ara\'ujo}\ \emph {et~al.}(2013)\citenamefont
		{Ara\'ujo}, \citenamefont {Quintino}, \citenamefont {Budroni}, \citenamefont
		{Cunha},\ and\ \citenamefont {Cabello}}]{nc_n_cycle}%
	\BibitemOpen
	\bibfield  {author} {\bibinfo {author} {\bibfnamefont {M.}~\bibnamefont
			{Ara\'ujo}}, \bibinfo {author} {\bibfnamefont {M.~T.}\ \bibnamefont
			{Quintino}}, \bibinfo {author} {\bibfnamefont {C.}~\bibnamefont {Budroni}},
		\bibinfo {author} {\bibfnamefont {M.~T.}\ \bibnamefont {Cunha}}, \ and\
		\bibinfo {author} {\bibfnamefont {A.}~\bibnamefont {Cabello}},\ }\href
	{\doibase 10.1103/PhysRevA.88.022118} {\bibfield  {journal} {\bibinfo
			{journal} {Phys. Rev. A}\ }\textbf {\bibinfo {volume} {88}},\ \bibinfo
		{pages} {022118} (\bibinfo {year} {2013})}\BibitemShut {NoStop}%
	\bibitem [{\citenamefont {Amaral}\ \emph {et~al.}(2014)\citenamefont {Amaral},
		\citenamefont {Cunha},\ and\ \citenamefont {Cabello}}]{e_principle_corr}%
	\BibitemOpen
	\bibfield  {author} {\bibinfo {author} {\bibfnamefont {B.}~\bibnamefont
			{Amaral}}, \bibinfo {author} {\bibfnamefont {M.~T.}\ \bibnamefont {Cunha}}, \
		and\ \bibinfo {author} {\bibfnamefont {A.}~\bibnamefont {Cabello}},\ }\href
	{\doibase 10.1103/PhysRevA.89.030101} {\bibfield  {journal} {\bibinfo
			{journal} {Phys. Rev. A}\ }\textbf {\bibinfo {volume} {89}},\ \bibinfo
		{pages} {030101} (\bibinfo {year} {2014})}\BibitemShut {NoStop}%
	\bibitem [{\citenamefont {Kurzy\ifmmode~\acute{n}\else \'{n}\fi{}ski}\ and\
		\citenamefont {Kaszlikowski}(2012)}]{context_dago}%
	\BibitemOpen
	\bibfield  {author} {\bibinfo {author} {\bibfnamefont {P.}~\bibnamefont
			{Kurzy\ifmmode~\acute{n}\else \'{n}\fi{}ski}}\ and\ \bibinfo {author}
		{\bibfnamefont {D.}~\bibnamefont {Kaszlikowski}},\ }\href {\doibase
		10.1103/PhysRevA.86.042125} {\bibfield  {journal} {\bibinfo  {journal} {Phys.
				Rev. A}\ }\textbf {\bibinfo {volume} {86}},\ \bibinfo {pages} {042125}
		(\bibinfo {year} {2012})}\BibitemShut {NoStop}%
	\bibitem [{\citenamefont {Lison\ifmmode~\check{e}\else \v{e}\fi{}k}\ \emph
		{et~al.}(2014)\citenamefont {Lison\ifmmode~\check{e}\else \v{e}\fi{}k},
		\citenamefont {Badzia\ifmmode~\mbox{\c{}}\else \c{}\fi{}g}, \citenamefont
		{Portillo},\ and\ \citenamefont {Cabello}}]{context_seven_set}%
	\BibitemOpen
	\bibfield  {author} {\bibinfo {author} {\bibfnamefont {P.}~\bibnamefont
			{Lison\ifmmode~\check{e}\else \v{e}\fi{}k}}, \bibinfo {author} {\bibfnamefont
			{P.}~\bibnamefont {Badzia\ifmmode~\mbox{\c{}}\else \c{}\fi{}g}}, \bibinfo
		{author} {\bibfnamefont {J.~R.}\ \bibnamefont {Portillo}}, \ and\ \bibinfo
		{author} {\bibfnamefont {A.}~\bibnamefont {Cabello}},\ }\href {\doibase
		10.1103/PhysRevA.89.042101} {\bibfield  {journal} {\bibinfo  {journal} {Phys.
				Rev. A}\ }\textbf {\bibinfo {volume} {89}},\ \bibinfo {pages} {042101}
		(\bibinfo {year} {2014})}\BibitemShut {NoStop}%
	\bibitem [{\citenamefont {Cabello}\ \emph {et~al.}(2015)\citenamefont
		{Cabello}, \citenamefont {Kleinmann},\ and\ \citenamefont
		{Budroni}}]{necessary_cond_context}%
	\BibitemOpen
	\bibfield  {author} {\bibinfo {author} {\bibfnamefont {A.}~\bibnamefont
			{Cabello}}, \bibinfo {author} {\bibfnamefont {M.}~\bibnamefont {Kleinmann}},
		\ and\ \bibinfo {author} {\bibfnamefont {C.}~\bibnamefont {Budroni}},\ }\href
	{\doibase 10.1103/PhysRevLett.114.250402} {\bibfield  {journal} {\bibinfo
			{journal} {Phys. Rev. Lett.}\ }\textbf {\bibinfo {volume} {114}},\ \bibinfo
		{pages} {250402} (\bibinfo {year} {2015})}\BibitemShut {NoStop}%
	\bibitem [{\citenamefont {Kleinmann}\ \emph {et~al.}(2012)\citenamefont
		{Kleinmann}, \citenamefont {Budroni}, \citenamefont {Larsson}, \citenamefont
		{G\"uhne},\ and\ \citenamefont {Cabello}}]{optimal_ineq_context}%
	\BibitemOpen
	\bibfield  {author} {\bibinfo {author} {\bibfnamefont {M.}~\bibnamefont
			{Kleinmann}}, \bibinfo {author} {\bibfnamefont {C.}~\bibnamefont {Budroni}},
		\bibinfo {author} {\bibfnamefont {J.-A.}\ \bibnamefont {Larsson}}, \bibinfo
		{author} {\bibfnamefont {O.}~\bibnamefont {G\"uhne}}, \ and\ \bibinfo
		{author} {\bibfnamefont {A.}~\bibnamefont {Cabello}},\ }\href {\doibase
		10.1103/PhysRevLett.109.250402} {\bibfield  {journal} {\bibinfo  {journal}
			{Phys. Rev. Lett.}\ }\textbf {\bibinfo {volume} {109}},\ \bibinfo {pages}
		{250402} (\bibinfo {year} {2012})}\BibitemShut {NoStop}%
	\bibitem [{\citenamefont {Cabello}\ \emph {et~al.}(2016)\citenamefont
		{Cabello}, \citenamefont {Kleinmann},\ and\ \citenamefont
		{Portillo}}]{cabello_13_ray}%
	\BibitemOpen
	\bibfield  {author} {\bibinfo {author} {\bibfnamefont {A.}~\bibnamefont
			{Cabello}}, \bibinfo {author} {\bibfnamefont {M.}~\bibnamefont {Kleinmann}},
		\ and\ \bibinfo {author} {\bibfnamefont {J.~R.}\ \bibnamefont {Portillo}},\
	}\href {\doibase 10.1088/1751-8113/49/38/38lt01} {\bibfield  {journal}
		{\bibinfo  {journal} {Journal of Physics A: Mathematical and Theoretical}\
		}\textbf {\bibinfo {volume} {49}},\ \bibinfo {pages} {38LT01} (\bibinfo
		{year} {2016})}\BibitemShut {NoStop}%
	\bibitem [{\citenamefont {Yu}\ and\ \citenamefont {Oh}(2012)}]{oh_13_ray}%
	\BibitemOpen
	\bibfield  {author} {\bibinfo {author} {\bibfnamefont {S.}~\bibnamefont
			{Yu}}\ and\ \bibinfo {author} {\bibfnamefont {C.~H.}\ \bibnamefont {Oh}},\
	}\href {\doibase 10.1103/PhysRevLett.108.030402} {\bibfield  {journal}
		{\bibinfo  {journal} {Phys. Rev. Lett.}\ }\textbf {\bibinfo {volume} {108}},\
		\bibinfo {pages} {030402} (\bibinfo {year} {2012})}\BibitemShut {NoStop}%
	\bibitem [{\citenamefont {Cabello}(2013)}]{cabello_twin_ineq}%
	\BibitemOpen
	\bibfield  {author} {\bibinfo {author} {\bibfnamefont {A.}~\bibnamefont
			{Cabello}},\ }\href {\doibase 10.1103/PhysRevA.87.010104} {\bibfield
		{journal} {\bibinfo  {journal} {Phys. Rev. A}\ }\textbf {\bibinfo {volume}
			{87}},\ \bibinfo {pages} {010104} (\bibinfo {year} {2013})}\BibitemShut
	{NoStop}%
	\bibitem [{\citenamefont {Bharti}\ \emph
		{et~al.}(2019{\natexlab{a}})\citenamefont {Bharti}, \citenamefont {Ray},\
		and\ \citenamefont {Kwek}}]{kishor_context_ncycle}%
	\BibitemOpen
	\bibfield  {author} {\bibinfo {author} {\bibfnamefont {K.}~\bibnamefont
			{Bharti}}, \bibinfo {author} {\bibfnamefont {M.}~\bibnamefont {Ray}}, \ and\
		\bibinfo {author} {\bibfnamefont {L.-C.}\ \bibnamefont {Kwek}},\ }\href
	{\doibase 10.3390/e21020134} {\bibfield  {journal} {\bibinfo  {journal}
			{Entropy}\ }\textbf {\bibinfo {volume} {21}} (\bibinfo {year}
		{2019}{\natexlab{a}}),\ 10.3390/e21020134}\BibitemShut {NoStop}%
	\bibitem [{\citenamefont {Ramanathan}\ \emph {et~al.}(2012)\citenamefont
		{Ramanathan}, \citenamefont {Soeda}, \citenamefont
		{Kurzy\ifmmode~\acute{n}\else \'{n}\fi{}ski},\ and\ \citenamefont
		{Kaszlikowski}}]{context_mono}%
	\BibitemOpen
	\bibfield  {author} {\bibinfo {author} {\bibfnamefont {R.}~\bibnamefont
			{Ramanathan}}, \bibinfo {author} {\bibfnamefont {A.}~\bibnamefont {Soeda}},
		\bibinfo {author} {\bibfnamefont {P.}~\bibnamefont
			{Kurzy\ifmmode~\acute{n}\else \'{n}\fi{}ski}}, \ and\ \bibinfo {author}
		{\bibfnamefont {D.}~\bibnamefont {Kaszlikowski}},\ }\href {\doibase
		10.1103/PhysRevLett.109.050404} {\bibfield  {journal} {\bibinfo  {journal}
			{Phys. Rev. Lett.}\ }\textbf {\bibinfo {volume} {109}},\ \bibinfo {pages}
		{050404} (\bibinfo {year} {2012})}\BibitemShut {NoStop}%
	\bibitem [{\citenamefont {Kurzy\ifmmode~\acute{n}\else \'{n}\fi{}ski}\ \emph
		{et~al.}(2014)\citenamefont {Kurzy\ifmmode~\acute{n}\else \'{n}\fi{}ski},
		\citenamefont {Cabello},\ and\ \citenamefont
		{Kaszlikowski}}]{context_nonlocal_mono}%
	\BibitemOpen
	\bibfield  {author} {\bibinfo {author} {\bibfnamefont {P.}~\bibnamefont
			{Kurzy\ifmmode~\acute{n}\else \'{n}\fi{}ski}}, \bibinfo {author}
		{\bibfnamefont {A.}~\bibnamefont {Cabello}}, \ and\ \bibinfo {author}
		{\bibfnamefont {D.}~\bibnamefont {Kaszlikowski}},\ }\href {\doibase
		10.1103/PhysRevLett.112.100401} {\bibfield  {journal} {\bibinfo  {journal}
			{Phys. Rev. Lett.}\ }\textbf {\bibinfo {volume} {112}},\ \bibinfo {pages}
		{100401} (\bibinfo {year} {2014})}\BibitemShut {NoStop}%
	\bibitem [{\citenamefont {Bharti}\ \emph
		{et~al.}(2019{\natexlab{b}})\citenamefont {Bharti}, \citenamefont {Ray},
		\citenamefont {Varvitsiotis}, \citenamefont {Warsi}, \citenamefont
		{Cabello},\ and\ \citenamefont {Kwek}}]{kishor_robust}%
	\BibitemOpen
	\bibfield  {author} {\bibinfo {author} {\bibfnamefont {K.}~\bibnamefont
			{Bharti}}, \bibinfo {author} {\bibfnamefont {M.}~\bibnamefont {Ray}},
		\bibinfo {author} {\bibfnamefont {A.}~\bibnamefont {Varvitsiotis}}, \bibinfo
		{author} {\bibfnamefont {N.~A.}\ \bibnamefont {Warsi}}, \bibinfo {author}
		{\bibfnamefont {A.}~\bibnamefont {Cabello}}, \ and\ \bibinfo {author}
		{\bibfnamefont {L.-C.}\ \bibnamefont {Kwek}},\ }\href {\doibase
		10.1103/PhysRevLett.122.250403} {\bibfield  {journal} {\bibinfo  {journal}
			{Phys. Rev. Lett.}\ }\textbf {\bibinfo {volume} {122}},\ \bibinfo {pages}
		{250403} (\bibinfo {year} {2019}{\natexlab{b}})}\BibitemShut {NoStop}%
	\bibitem [{\citenamefont {Singh}\ \emph {et~al.}(2017)\citenamefont {Singh},
		\citenamefont {Bharti},\ and\ \citenamefont {Arvind}}]{jask_mono}%
	\BibitemOpen
	\bibfield  {author} {\bibinfo {author} {\bibfnamefont {J.}~\bibnamefont
			{Singh}}, \bibinfo {author} {\bibfnamefont {K.}~\bibnamefont {Bharti}}, \
		and\ \bibinfo {author} {\bibnamefont {Arvind}},\ }\href {\doibase
		10.1103/PhysRevA.95.062333} {\bibfield  {journal} {\bibinfo  {journal} {Phys.
				Rev. A}\ }\textbf {\bibinfo {volume} {95}},\ \bibinfo {pages} {062333}
		(\bibinfo {year} {2017})}\BibitemShut {NoStop}%
	\bibitem [{\citenamefont {Cabello}\ \emph {et~al.}(2011)\citenamefont
		{Cabello}, \citenamefont {D'Ambrosio}, \citenamefont {Nagali},\ and\
		\citenamefont {Sciarrino}}]{cabello_context_keydist}%
	\BibitemOpen
	\bibfield  {author} {\bibinfo {author} {\bibfnamefont {A.}~\bibnamefont
			{Cabello}}, \bibinfo {author} {\bibfnamefont {V.}~\bibnamefont {D'Ambrosio}},
		\bibinfo {author} {\bibfnamefont {E.}~\bibnamefont {Nagali}}, \ and\ \bibinfo
		{author} {\bibfnamefont {F.}~\bibnamefont {Sciarrino}},\ }\href {\doibase
		10.1103/PhysRevA.84.030302} {\bibfield  {journal} {\bibinfo  {journal} {Phys.
				Rev. A}\ }\textbf {\bibinfo {volume} {84}},\ \bibinfo {pages} {030302}
		(\bibinfo {year} {2011})}\BibitemShut {NoStop}%
	\bibitem [{\citenamefont {Troupe}\ and\ \citenamefont
		{Farinholt}(2015)}]{troupe_context_keydist}%
	\BibitemOpen
	\bibfield  {author} {\bibinfo {author} {\bibfnamefont {J.~E.}\ \bibnamefont
			{Troupe}}\ and\ \bibinfo {author} {\bibfnamefont {J.~M.}\ \bibnamefont
			{Farinholt}},\ }\href@noop {} {\  (\bibinfo {year} {2015})},\ \Eprint
	{http://arxiv.org/abs/1512.02256} {arXiv:1512.02256 [quant-ph]} \BibitemShut
	{NoStop}%
	\bibitem [{\citenamefont {Saha}\ \emph {et~al.}(2019)\citenamefont {Saha},
		\citenamefont {Horodecki},\ and\ \citenamefont {Paw{\l}owski}}]{debasish2}%
	\BibitemOpen
	\bibfield  {author} {\bibinfo {author} {\bibfnamefont {D.}~\bibnamefont
			{Saha}}, \bibinfo {author} {\bibfnamefont {P.}~\bibnamefont {Horodecki}}, \
		and\ \bibinfo {author} {\bibfnamefont {M.}~\bibnamefont {Paw{\l}owski}},\
	}\href {\doibase 10.1088/1367-2630/ab4149} {\bibfield  {journal} {\bibinfo
			{journal} {New Journal of Physics}\ }\textbf {\bibinfo {volume} {21}},\
		\bibinfo {pages} {093057} (\bibinfo {year} {2019})}\BibitemShut {NoStop}%
	\bibitem [{\citenamefont {Spekkens}(2005)}]{spekkens_context}%
	\BibitemOpen
	\bibfield  {author} {\bibinfo {author} {\bibfnamefont {R.~W.}\ \bibnamefont
			{Spekkens}},\ }\href {\doibase 10.1103/PhysRevA.71.052108} {\bibfield
		{journal} {\bibinfo  {journal} {Phys. Rev. A}\ }\textbf {\bibinfo {volume}
			{71}},\ \bibinfo {pages} {052108} (\bibinfo {year} {2005})}\BibitemShut
	{NoStop}%
	\bibitem [{\citenamefont {Spekkens}(2014)}]{spekkens1}%
	\BibitemOpen
	\bibfield  {author} {\bibinfo {author} {\bibfnamefont {R.~W.}\ \bibnamefont
			{Spekkens}},\ }\href {\doibase 10.1007/s10701-014-9833-x} {\bibfield
		{journal} {\bibinfo  {journal} {Foundations of Physics}\ }\textbf {\bibinfo
			{volume} {44}},\ \bibinfo {pages} {1125} (\bibinfo {year}
		{2014})}\BibitemShut {NoStop}%
	\bibitem [{\citenamefont {Schmid}\ \emph {et~al.}(2018)\citenamefont {Schmid},
		\citenamefont {Spekkens},\ and\ \citenamefont {Wolfe}}]{schmid3}%
	\BibitemOpen
	\bibfield  {author} {\bibinfo {author} {\bibfnamefont {D.}~\bibnamefont
			{Schmid}}, \bibinfo {author} {\bibfnamefont {R.~W.}\ \bibnamefont
			{Spekkens}}, \ and\ \bibinfo {author} {\bibfnamefont {E.}~\bibnamefont
			{Wolfe}},\ }\href {\doibase 10.1103/PhysRevA.97.062103} {\bibfield  {journal}
		{\bibinfo  {journal} {Phys. Rev. A}\ }\textbf {\bibinfo {volume} {97}},\
		\bibinfo {pages} {062103} (\bibinfo {year} {2018})}\BibitemShut {NoStop}%
	\bibitem [{\citenamefont {Spekkens}\ \emph
		{et~al.}(2009{\natexlab{a}})\citenamefont {Spekkens}, \citenamefont
		{Buzacott}, \citenamefont {Keehn}, \citenamefont {Toner},\ and\ \citenamefont
		{Pryde}}]{spekkens_parity_context}%
	\BibitemOpen
	\bibfield  {author} {\bibinfo {author} {\bibfnamefont {R.~W.}\ \bibnamefont
			{Spekkens}}, \bibinfo {author} {\bibfnamefont {D.~H.}\ \bibnamefont
			{Buzacott}}, \bibinfo {author} {\bibfnamefont {A.~J.}\ \bibnamefont {Keehn}},
		\bibinfo {author} {\bibfnamefont {B.}~\bibnamefont {Toner}}, \ and\ \bibinfo
		{author} {\bibfnamefont {G.~J.}\ \bibnamefont {Pryde}},\ }\href {\doibase
		10.1103/PhysRevLett.102.010401} {\bibfield  {journal} {\bibinfo  {journal}
			{Phys. Rev. Lett.}\ }\textbf {\bibinfo {volume} {102}},\ \bibinfo {pages}
		{010401} (\bibinfo {year} {2009}{\natexlab{a}})}\BibitemShut {NoStop}%
	\bibitem [{\citenamefont {Tavakoli}\ \emph {et~al.}(2021)\citenamefont
		{Tavakoli}, \citenamefont {Cruzeiro}, \citenamefont {Uola},\ and\
		\citenamefont {Abbott}}]{context_corr_quantum}%
	\BibitemOpen
	\bibfield  {author} {\bibinfo {author} {\bibfnamefont {A.}~\bibnamefont
			{Tavakoli}}, \bibinfo {author} {\bibfnamefont {E.~Z.}\ \bibnamefont
			{Cruzeiro}}, \bibinfo {author} {\bibfnamefont {R.}~\bibnamefont {Uola}}, \
		and\ \bibinfo {author} {\bibfnamefont {A.~A.}\ \bibnamefont {Abbott}},\
	}\href {\doibase 10.1103/PRXQuantum.2.020334} {\bibfield  {journal} {\bibinfo
			{journal} {PRX Quantum}\ }\textbf {\bibinfo {volume} {2}},\ \bibinfo {pages}
		{020334} (\bibinfo {year} {2021})}\BibitemShut {NoStop}%
	\bibitem [{\citenamefont {Ambainis}\ \emph
		{et~al.}(2019{\natexlab{a}})\citenamefont {Ambainis}, \citenamefont {Banik},
		\citenamefont {Chaturvedi}, \citenamefont {Kravchenko},\ and\ \citenamefont
		{Rai}}]{rac_context}%
	\BibitemOpen
	\bibfield  {author} {\bibinfo {author} {\bibfnamefont {A.}~\bibnamefont
			{Ambainis}}, \bibinfo {author} {\bibfnamefont {M.}~\bibnamefont {Banik}},
		\bibinfo {author} {\bibfnamefont {A.}~\bibnamefont {Chaturvedi}}, \bibinfo
		{author} {\bibfnamefont {D.}~\bibnamefont {Kravchenko}}, \ and\ \bibinfo
		{author} {\bibfnamefont {A.}~\bibnamefont {Rai}},\ }\href {\doibase
		10.1007/s11128-019-2228-3} {\bibfield  {journal} {\bibinfo  {journal}
			{Quantum Information Processing}\ }\textbf {\bibinfo {volume} {18}},\
		\bibinfo {pages} {111} (\bibinfo {year} {2019}{\natexlab{a}})}\BibitemShut
	{NoStop}%
	\bibitem [{\citenamefont {Schmid}\ and\ \citenamefont
		{Spekkens}(2018)}]{schmid_state_discrimination_context}%
	\BibitemOpen
	\bibfield  {author} {\bibinfo {author} {\bibfnamefont {D.}~\bibnamefont
			{Schmid}}\ and\ \bibinfo {author} {\bibfnamefont {R.~W.}\ \bibnamefont
			{Spekkens}},\ }\href {\doibase 10.1103/PhysRevX.8.011015} {\bibfield
		{journal} {\bibinfo  {journal} {Phys. Rev. X}\ }\textbf {\bibinfo {volume}
			{8}},\ \bibinfo {pages} {011015} (\bibinfo {year} {2018})}\BibitemShut
	{NoStop}%
	\bibitem [{\citenamefont {Chaturvedi}\ \emph {et~al.}(2021)\citenamefont
		{Chaturvedi}, \citenamefont {Farkas},\ and\ \citenamefont
		{Wright}}]{anubhav}%
	\BibitemOpen
	\bibfield  {author} {\bibinfo {author} {\bibfnamefont {A.}~\bibnamefont
			{Chaturvedi}}, \bibinfo {author} {\bibfnamefont {M.}~\bibnamefont {Farkas}},
		\ and\ \bibinfo {author} {\bibfnamefont {V.~J.}\ \bibnamefont {Wright}},\
	}\href {\doibase 10.22331/q-2021-06-29-484} {\bibfield  {journal} {\bibinfo
			{journal} {{Quantum}}\ }\textbf {\bibinfo {volume} {5}},\ \bibinfo {pages}
		{484} (\bibinfo {year} {2021})}\BibitemShut {NoStop}%
	\bibitem [{\citenamefont {Pusey}(2014)}]{pusey}%
	\BibitemOpen
	\bibfield  {author} {\bibinfo {author} {\bibfnamefont {M.~F.}\ \bibnamefont
			{Pusey}},\ }\href {\doibase 10.1103/PhysRevLett.113.200401} {\bibfield
		{journal} {\bibinfo  {journal} {Phys. Rev. Lett.}\ }\textbf {\bibinfo
			{volume} {113}},\ \bibinfo {pages} {200401} (\bibinfo {year}
		{2014})}\BibitemShut {NoStop}%
	\bibitem [{\citenamefont {Pusey}\ and\ \citenamefont {Leifer}(2015)}]{pusey2}%
	\BibitemOpen
	\bibfield  {author} {\bibinfo {author} {\bibfnamefont {M.~F.}\ \bibnamefont
			{Pusey}}\ and\ \bibinfo {author} {\bibfnamefont {M.~S.}\ \bibnamefont
			{Leifer}},\ }\href {\doibase 10.4204/eptcs.195.22} {\bibfield  {journal}
		{\bibinfo  {journal} {Electronic Proceedings in Theoretical Computer
				Science}\ }\textbf {\bibinfo {volume} {195}},\ \bibinfo {pages} {295–306}
		(\bibinfo {year} {2015})}\BibitemShut {NoStop}%
	\bibitem [{\citenamefont {Mazurek}\ \emph {et~al.}(2016)\citenamefont
		{Mazurek}, \citenamefont {Pusey}, \citenamefont {Kunjwal}, \citenamefont
		{Resch},\ and\ \citenamefont {Spekkens}}]{kunjwal_noncontext_unphys}%
	\BibitemOpen
	\bibfield  {author} {\bibinfo {author} {\bibfnamefont {M.~D.}\ \bibnamefont
			{Mazurek}}, \bibinfo {author} {\bibfnamefont {M.~F.}\ \bibnamefont {Pusey}},
		\bibinfo {author} {\bibfnamefont {R.}~\bibnamefont {Kunjwal}}, \bibinfo
		{author} {\bibfnamefont {K.~J.}\ \bibnamefont {Resch}}, \ and\ \bibinfo
		{author} {\bibfnamefont {R.~W.}\ \bibnamefont {Spekkens}},\ }\href
	{https://doi.org/10.1038/ncomms11780} {\bibfield  {journal} {\bibinfo
			{journal} {Nature Communications}\ }\textbf {\bibinfo {volume} {7}},\
		\bibinfo {pages} {ncomms11780 EP } (\bibinfo {year} {2016})}\BibitemShut
	{NoStop}%
	\bibitem [{\citenamefont {Kunjwal}\ and\ \citenamefont
		{Spekkens}(2018)}]{kunjwal_stat_robust_context}%
	\BibitemOpen
	\bibfield  {author} {\bibinfo {author} {\bibfnamefont {R.}~\bibnamefont
			{Kunjwal}}\ and\ \bibinfo {author} {\bibfnamefont {R.~W.}\ \bibnamefont
			{Spekkens}},\ }\href {\doibase 10.1103/PhysRevA.97.052110} {\bibfield
		{journal} {\bibinfo  {journal} {Phys. Rev. A}\ }\textbf {\bibinfo {volume}
			{97}},\ \bibinfo {pages} {052110} (\bibinfo {year} {2018})}\BibitemShut
	{NoStop}%
	\bibitem [{\citenamefont {Kunjwal}\ and\ \citenamefont
		{Spekkens}(2015)}]{kunjwal_kstheorem_no_det}%
	\BibitemOpen
	\bibfield  {author} {\bibinfo {author} {\bibfnamefont {R.}~\bibnamefont
			{Kunjwal}}\ and\ \bibinfo {author} {\bibfnamefont {R.~W.}\ \bibnamefont
			{Spekkens}},\ }\href {\doibase 10.1103/PhysRevLett.115.110403} {\bibfield
		{journal} {\bibinfo  {journal} {Phys. Rev. Lett.}\ }\textbf {\bibinfo
			{volume} {115}},\ \bibinfo {pages} {110403} (\bibinfo {year}
		{2015})}\BibitemShut {NoStop}%
	\bibitem [{\citenamefont {Xu}\ \emph {et~al.}(2016)\citenamefont {Xu},
		\citenamefont {Saha}, \citenamefont {Su}, \citenamefont {Paw\l{}owski},\ and\
		\citenamefont {Chen}}]{debasish1}%
	\BibitemOpen
	\bibfield  {author} {\bibinfo {author} {\bibfnamefont {Z.-P.}\ \bibnamefont
			{Xu}}, \bibinfo {author} {\bibfnamefont {D.}~\bibnamefont {Saha}}, \bibinfo
		{author} {\bibfnamefont {H.-Y.}\ \bibnamefont {Su}}, \bibinfo {author}
		{\bibfnamefont {M.}~\bibnamefont {Paw\l{}owski}}, \ and\ \bibinfo {author}
		{\bibfnamefont {J.-L.}\ \bibnamefont {Chen}},\ }\href {\doibase
		10.1103/PhysRevA.94.062103} {\bibfield  {journal} {\bibinfo  {journal} {Phys.
				Rev. A}\ }\textbf {\bibinfo {volume} {94}},\ \bibinfo {pages} {062103}
		(\bibinfo {year} {2016})}\BibitemShut {NoStop}%
	\bibitem [{\citenamefont {Anwer}\ \emph {et~al.}(2021)\citenamefont {Anwer},
		\citenamefont {Wilson}, \citenamefont {Silva}, \citenamefont {Muhammad},
		\citenamefont {Tavakoli},\ and\ \citenamefont {Bourennane}}]{anwer}%
	\BibitemOpen
	\bibfield  {author} {\bibinfo {author} {\bibfnamefont {H.}~\bibnamefont
			{Anwer}}, \bibinfo {author} {\bibfnamefont {N.}~\bibnamefont {Wilson}},
		\bibinfo {author} {\bibfnamefont {R.}~\bibnamefont {Silva}}, \bibinfo
		{author} {\bibfnamefont {S.}~\bibnamefont {Muhammad}}, \bibinfo {author}
		{\bibfnamefont {A.}~\bibnamefont {Tavakoli}}, \ and\ \bibinfo {author}
		{\bibfnamefont {M.}~\bibnamefont {Bourennane}},\ }\href {\doibase
		10.22331/q-2021-09-28-551} {\bibfield  {journal} {\bibinfo  {journal}
			{{Quantum}}\ }\textbf {\bibinfo {volume} {5}},\ \bibinfo {pages} {551}
		(\bibinfo {year} {2021})}\BibitemShut {NoStop}%
	\bibitem [{\citenamefont {Spekkens}\ \emph
		{et~al.}(2009{\natexlab{b}})\citenamefont {Spekkens}, \citenamefont
		{Buzacott}, \citenamefont {Keehn}, \citenamefont {Toner},\ and\ \citenamefont
		{Pryde}}]{spekkens_prep_context_app}%
	\BibitemOpen
	\bibfield  {author} {\bibinfo {author} {\bibfnamefont {R.~W.}\ \bibnamefont
			{Spekkens}}, \bibinfo {author} {\bibfnamefont {D.~H.}\ \bibnamefont
			{Buzacott}}, \bibinfo {author} {\bibfnamefont {A.~J.}\ \bibnamefont {Keehn}},
		\bibinfo {author} {\bibfnamefont {B.}~\bibnamefont {Toner}}, \ and\ \bibinfo
		{author} {\bibfnamefont {G.~J.}\ \bibnamefont {Pryde}},\ }\href {\doibase
		10.1103/PhysRevLett.102.010401} {\bibfield  {journal} {\bibinfo  {journal}
			{Phys. Rev. Lett.}\ }\textbf {\bibinfo {volume} {102}},\ \bibinfo {pages}
		{010401} (\bibinfo {year} {2009}{\natexlab{b}})}\BibitemShut {NoStop}%
	\bibitem [{\citenamefont {Chailloux}\ \emph {et~al.}(2016)\citenamefont
		{Chailloux}, \citenamefont {Kerenidis}, \citenamefont {Kundu},\ and\
		\citenamefont {Sikora}}]{sikora_random_access_codes}%
	\BibitemOpen
	\bibfield  {author} {\bibinfo {author} {\bibfnamefont {A.}~\bibnamefont
			{Chailloux}}, \bibinfo {author} {\bibfnamefont {I.}~\bibnamefont
			{Kerenidis}}, \bibinfo {author} {\bibfnamefont {S.}~\bibnamefont {Kundu}}, \
		and\ \bibinfo {author} {\bibfnamefont {J.}~\bibnamefont {Sikora}},\ }\href
	{\doibase 10.1088/1367-2630/18/4/045003} {\bibfield  {journal} {\bibinfo
			{journal} {New Journal of Physics}\ }\textbf {\bibinfo {volume} {18}},\
		\bibinfo {pages} {045003} (\bibinfo {year} {2016})}\BibitemShut {NoStop}%
	\bibitem [{\citenamefont {Ambainis}\ \emph
		{et~al.}(2019{\natexlab{b}})\citenamefont {Ambainis}, \citenamefont {Banik},
		\citenamefont {Chaturvedi}, \citenamefont {Kravchenko},\ and\ \citenamefont
		{Rai}}]{banik_context_app}%
	\BibitemOpen
	\bibfield  {author} {\bibinfo {author} {\bibfnamefont {A.}~\bibnamefont
			{Ambainis}}, \bibinfo {author} {\bibfnamefont {M.}~\bibnamefont {Banik}},
		\bibinfo {author} {\bibfnamefont {A.}~\bibnamefont {Chaturvedi}}, \bibinfo
		{author} {\bibfnamefont {D.}~\bibnamefont {Kravchenko}}, \ and\ \bibinfo
		{author} {\bibfnamefont {A.}~\bibnamefont {Rai}},\ }\href {\doibase
		10.1007/s11128-019-2228-3} {\bibfield  {journal} {\bibinfo  {journal}
			{Quantum Information Processing}\ }\textbf {\bibinfo {volume} {18}},\
		\bibinfo {pages} {111} (\bibinfo {year} {2019}{\natexlab{b}})}\BibitemShut
	{NoStop}%
	\bibitem [{\citenamefont {Kunjwal}\ and\ \citenamefont
		{Ghosh}(2014)}]{kunjwal_context_qubit}%
	\BibitemOpen
	\bibfield  {author} {\bibinfo {author} {\bibfnamefont {R.}~\bibnamefont
			{Kunjwal}}\ and\ \bibinfo {author} {\bibfnamefont {S.}~\bibnamefont
			{Ghosh}},\ }\href {\doibase 10.1103/PhysRevA.89.042118} {\bibfield  {journal}
		{\bibinfo  {journal} {Phys. Rev. A}\ }\textbf {\bibinfo {volume} {89}},\
		\bibinfo {pages} {042118} (\bibinfo {year} {2014})}\BibitemShut {NoStop}%
	\bibitem [{\citenamefont {Chaturvedi}\ and\ \citenamefont
		{Saha}(2020)}]{debasish4}%
	\BibitemOpen
	\bibfield  {author} {\bibinfo {author} {\bibfnamefont {A.}~\bibnamefont
			{Chaturvedi}}\ and\ \bibinfo {author} {\bibfnamefont {D.}~\bibnamefont
			{Saha}},\ }\href {\doibase 10.22331/q-2020-10-21-345} {\bibfield  {journal}
		{\bibinfo  {journal} {Quantum}\ }\textbf {\bibinfo {volume} {4}},\ \bibinfo
		{pages} {345} (\bibinfo {year} {2020})}\BibitemShut {NoStop}%
	\bibitem [{\citenamefont {Schmid}\ \emph {et~al.}(2020)\citenamefont {Schmid},
		\citenamefont {Selby},\ and\ \citenamefont {Spekkens}}]{schmid}%
	\BibitemOpen
	\bibfield  {author} {\bibinfo {author} {\bibfnamefont {D.}~\bibnamefont
			{Schmid}}, \bibinfo {author} {\bibfnamefont {J.~H.}\ \bibnamefont {Selby}}, \
		and\ \bibinfo {author} {\bibfnamefont {R.~W.}\ \bibnamefont {Spekkens}},\
	}\href@noop {} {\  (\bibinfo {year} {2020})},\ \Eprint
	{http://arxiv.org/abs/2009.03297} {arXiv:2009.03297 [quant-ph]} \BibitemShut
	{NoStop}%
	\bibitem [{\citenamefont {Selby}\ \emph {et~al.}(2021)\citenamefont {Selby},
		\citenamefont {Schmid}, \citenamefont {Wolfe}, \citenamefont {Sainz},
		\citenamefont {Kunjwal},\ and\ \citenamefont {Spekkens}}]{selby_context}%
	\BibitemOpen
	\bibfield  {author} {\bibinfo {author} {\bibfnamefont {J.~H.}\ \bibnamefont
			{Selby}}, \bibinfo {author} {\bibfnamefont {D.}~\bibnamefont {Schmid}},
		\bibinfo {author} {\bibfnamefont {E.}~\bibnamefont {Wolfe}}, \bibinfo
		{author} {\bibfnamefont {A.~B.}\ \bibnamefont {Sainz}}, \bibinfo {author}
		{\bibfnamefont {R.}~\bibnamefont {Kunjwal}}, \ and\ \bibinfo {author}
		{\bibfnamefont {R.~W.}\ \bibnamefont {Spekkens}},\ }\href@noop {} {\enquote
		{\bibinfo {title} {Contextuality without incompatibility},}\ } (\bibinfo
	{year} {2021}),\ \Eprint {http://arxiv.org/abs/2106.09045} {arXiv:2106.09045
		[quant-ph]} \BibitemShut {NoStop}%
\end{thebibliography}

%

\end{document}